\documentclass{appolb}
\usepackage{epsfig}
\begin{document}
\title{Confining but chirally symmetric quarkyonic matter%
\thanks{Presented at "Excited QCD 2011", 20-25 February, Les Houches, France}%
}
\author{L. Ya. Glozman
\address{Institute for Physics, Theoretical Physics Branch,
University of Graz, Universit\"atsplatz 5, A-8010 Graz, Austria}
}

\maketitle
\begin{abstract}
Here we overview a possible mechanism for confining but
chirally symmetric matter at low temperatures and large densities.
\end{abstract}
  
\section{Introduction}
One of the most intriguing questions of the strong interaction
physics is a possibility for existence of the chirally symmetric
matter with confinement at low temperatures and large densities.
It is quite possible that a confining mode persists up to a very
dense nuclear matter. Here, by a matter with confinement, we imply 
a matter with only the color singlet excitation modes of hadronic
type, where  uncorrelated colored excitation are not possible.
In the large $N_c$ limit confinement survives in a matter at low
temperatures up to arbitrary large density, because the quark - antiquark
loops as well as the quark - quark hole loops are suppressed at
large $N_c$. Nothing can screen a confining gluonic field and a gluodynamics
in a medium is the same as in a vacuum. In this case it is possible
to define a quarkyonic matter as a very dense nuclear matter where
some bulk properties, e.g.,  pressure, are determined by the quark Fermi surface
(like for a Fermi gas), while uncorrelated single quark excitations are 
not allowed \cite{pisarski}.  We do not know how to rigorously define
a confining mode in a matter at $N_c=3$, because there is no strict
order parameter for confinement. So we will imply under
 confining matter a matter with the color singlet excitation modes.
 We will also assume that such a confining matter exists in the
 real $N_c=3$ world up to  very large densities.\footnote{It is shown
 on the lattice for the $N_c=2$ QCD that at low temperatures
 confinement persists up
 to densities of the order 100 times nuclear matter density \cite{Hands}.
 It is quite natural then to assume that for the $N_c=3$ QCD confinement
 will survive even for essentially larger densities.}

What happens with chiral symmetry breaking
in a very dense cold matter with confinement? Is it
possible to have a chiral symmetry restoration phase
transition in a mode with confinement both below and
above the phase transition?

If the chiral restoration phase transition exists in a
system with confinement, then the origin of mass of the
confining strongly interacting matter above the phase 
transition is not related at all with the chiral symmetry
breaking. Such a possibility was not considered in the past
on the apriori grounds: It was believed that a hadron mass generation
is necessarily connected with the chiral symmetry breaking
in a vacuum and that a hadron mass in the light quark sector
is at least mostly directly related to the quark condensate of a vacuum.
Indeed, the 't Hooft anomaly matching conditions \cite{anomaly}
require that at zero density and temperature in a confining mode 
there must appear a Goldstone mode related to breaking of chiral
symmetry. However, the anomaly matching conditions can be trivially
satisfied in a two flavor nuclear matter built with chirally
symmetric baryons (i.e., without any Goldstone mode associated with spontaneous
breaking of chiral symmetry).  The Casher's
argument, claiming that in a confining mode the quark Green function
must necessarily contain a chiral symmetry breaking part \cite{Casher}
is not general enough and can be easily bypassed \cite{G1}. At last,
the effective restoration of chiral symmetry in highly excited hadrons
\cite{G2}, if finally established, will imply that the mass generation
mechanism of these hadrons is not related to the chiral symmetry
breaking in a vacuum. Summarizing, today there are no apriori arguments
that would rule out a possibility for a confining but chirally
symmetric dense and cold strongly interacting matter.

\section{Confining but chirally symmetric liquid phase}

In the large $N_c$ limit nucleons are infinitely heavy, 
a nuclear matter is in a crystal phase where  translational
and rotational invariances are spontaneously broken. In this
situation all possible chiral order parameters cannot 
simultaneously average to zero and if chiral symmetry is broken
locally, it is also broken in average \cite{cohen}. However, we
do know that in the real $N_c=3$ world a nuclear matter is a liquid
with manifest translational and rotational invariances. Then, it is
not unreasonable to assume
that a dense (and a superdense) $N_c=3$ baryonic matter with confinement
(i.e., by definition a quarkyonic matter) is also in a liquid phase.
Given this assumption, one can ask a question whether a chiral restoration
phase transition is possible or not in such a matter. If possible,
then by what mechanism?

We cannot answer this question from first principles. What
can be done, however, is to construct a model. If demonstrated
within such a model, this scenario could also be realized in QCD
and further theoretical efforts to clarify this interesting
questions would be called for. A minimal set of requirements for
such a model is that it must be manifestly confining, chirally
symmetric and  provide dynamical breaking of chiral symmetry in a vacuum.
This model must admit
solutions for hadrons with  nonzero mass as bound states of quarks
 both in the Wigner-Weyl and Nambu-Goldstone modes of chiral
symmetry. Such  program has been performed in refs. \cite{GW1,GW2,G1},
where it was explicitly demonstrated that a confining chirally symmetric
liquid phase at low temperature can indeed be obtained at least 
within a model that meets all requirements above.

\section{The model}

It is assumed within the model that the only interquark interaction
is a linear instantaneous potential of Coulomb type. Then the 
 $SU(2)_L \times SU(2)_R  \times U(1)_A \times U(1)_V $ 
 symmetric Hamiltonian is

\begin{eqnarray} 
\hat{H} & = & \int d^3x\bar{\psi}(\vec{x},t)\left(-i\vec{\gamma}\cdot
\vec{\bigtriangledown} \right)\psi(\vec{x},t) \nonumber \\
 &+& \frac12\int d^3
xd^3y\;J^a_\mu(\vec{x},t)K^{ab}_{\mu\nu}(\vec{x}-\vec{y})J^b_\nu(\vec{y},t),
\label{H} 
\end{eqnarray}

\noindent
 where  
 $J_{\mu}^a(\vec{x},t)=\bar{\psi}(\vec{x},t)\gamma_\mu\frac{\lambda^a}{2}
\psi(\vec{x},t)$ and the interaction is assumed to be

\begin{equation} 
K^{ab}_{\mu\nu}(\vec{x}-\vec{y})=g_{\mu 0}g_{\nu 0}
\delta^{ab} V (|\vec{x}-\vec{y}|); ~~~~~
\frac{\lambda^a \lambda^a}{4}V(r) = \sigma r,
\label{KK}
\end{equation}

\noindent
with $a,b$ being color indices.
This model was intensively used in the past to study chiral symmetry
breaking, chiral properties of hadrons, etc, \cite{Y}.
The model can be considered as a straightforward 3+1 dim
generalization of
the 1+1 dim 't Hooft model \cite{H}.
An important aspect of this 3+1 dim model is that it manifestly 
exhibits effective restoration of chiral symmetry in hadrons with
large spin $J$ \cite{largeJ}.

The self-energy of quarks in a vacuum,

\begin{equation}
\Sigma(\vec p) =A_p+(\vec{\gamma}\hat{\vec{p}})(B_p-p),
\label{SE} 
\end{equation}

\noindent
consists of the Lorentz-scalar chiral symmetry breaking part
$A_p$ and the chirally symmetric part $(\vec{\gamma}\hat{\vec{p}})(B_p-p)$.
The unknown functions $A_p$ and $B_p$ can be obtained from the
gap equation. All  color singlet quantities, like quark condensate,
hadron masses are finite and well defined. In contrast, all 
color non-singlet quantities, like a single quark energy, are
infinite. It is a manifestation of confinement.
 
We want to address chiral symmetry and confining properties
of a dense matter at $T=0$. We treat the system in a mean field
approximation and assume a simple valence quark distribution function,
see Fig. 1. In reality valence quarks near the Fermi surface interact
and cluster into the color singlet baryons. Then a rigid quark Fermi 
surface on Fig. 1 is diffused. Here we ignore these effects.\footnote{A
reasonable diffusion of the quark Fermi surface does not lead to
qualitative modifications of results \cite{GSW}.}

\begin{figure}[h]
  \begin{center}
    
      \center{\includegraphics[width=0.3\linewidth]{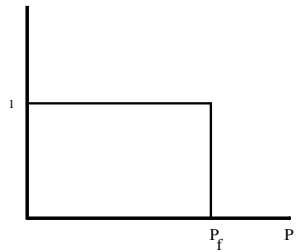} \\ }
    \end{center}
  \caption{Valence quark distribution.}
\end{figure}

In a dense matter at $T=0$ the most important physics that leads
to  restoration of chiral symmetry is the Pauli blocking by
valence quarks of the positive energy levels required for the very
existence of the quark condensate. This is similar to the chiral
restoration in the Nambu and Jona-Lasinio model \cite{NJL}.
At sufficiently large Fermi momentum the gap equation does not
admit a nontrivial solution with broken chiral symmetry. Consequently,
the chiral symmetry breaking Lorentz scalar part $A_p$ of the quark
self-energy vanishes and the chiral symmetry gets restored, see Fig. 2. 
However,
the chirally symmetric part of the quark self-energy does not vanish
and is still infrared divergent, like in vacuum. This means that even with 
restored chiral
symmetry a single quark energy is infinite and such a quark is confined.
This infrared divergence cancels exactly in all color singlet hadronic
modes that remain finite and well defined.

\begin{figure}
\begin{center}
 \includegraphics[width=0.5\hsize,clip=]{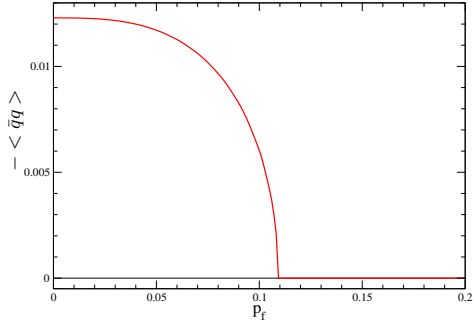}
 \caption{Quark condensate in units of $\sigma^{3/2}$
 as a function of the Fermi momentum, which is units of $\sqrt \sigma$.}
\end{center}
 \end{figure}

In this respect the model is radically different from the non-confining
NJL model. In the latter a dense matter is a Fermi gas of free
quarks. In the Nambu - Goldstone mode these quarks are massive.
In the Wigner-Weyl mode they are massless. In our case
physical degrees of freedom, that can be excited, 
are color singlet hadrons. In the
Wigner-Weyl mode these are the chirally symmetric hadrons.

Given a quark Green function  from the gap equation,
one can solve the Bethe-Salpeter equation to obtain the color
singlet mesons. A spectrum of all possible quark-antiquark mesons
below and above the chiral restoration phase transition is shown
on Fig. 3 and 4. Obviously, the chiral symmetry is manifestly
broken on Fig. 3, while on Fig. 4 the hadrons fall into all
possible chiral multiplets. Technically it is more difficult
to solve the model for baryons, but in principle it can be done.

\begin{figure*}
\mbox{
\includegraphics[width=0.16\hsize]{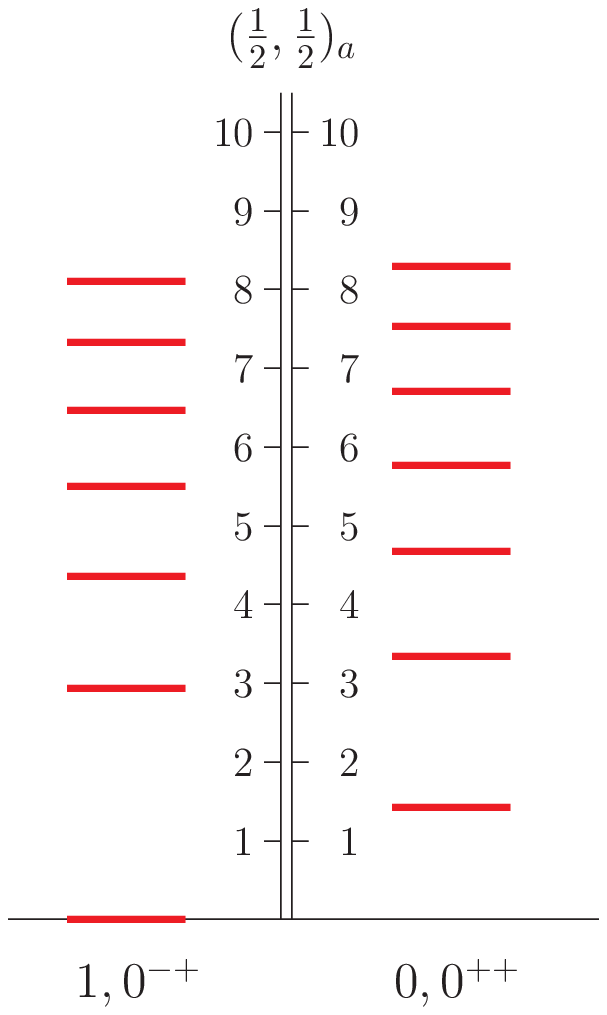}\,%
\includegraphics[width=0.16\hsize]{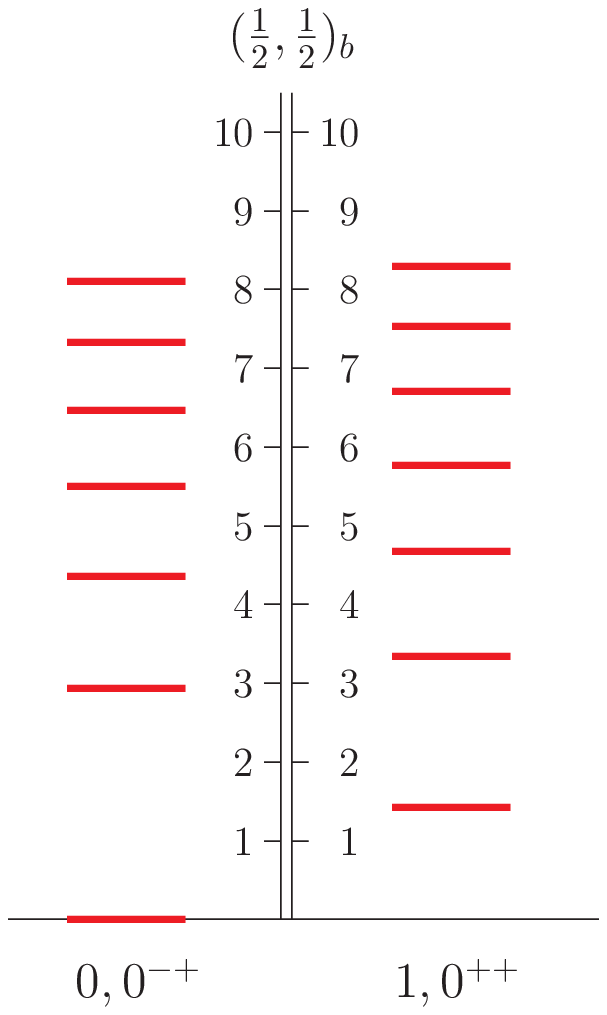}\,%
\includegraphics[width=0.16\hsize]{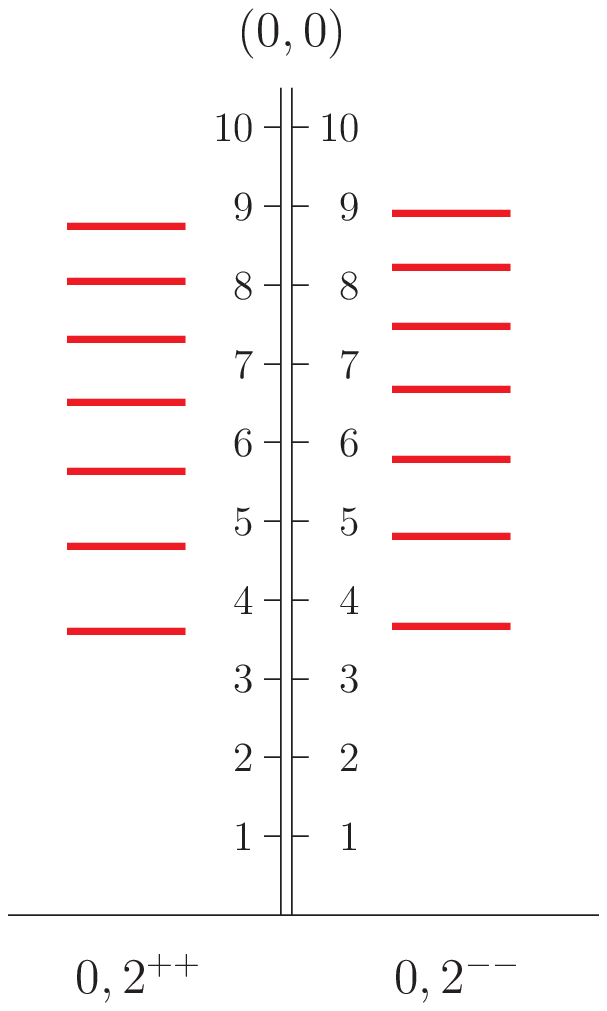}\,%
\includegraphics[width=0.16\hsize]{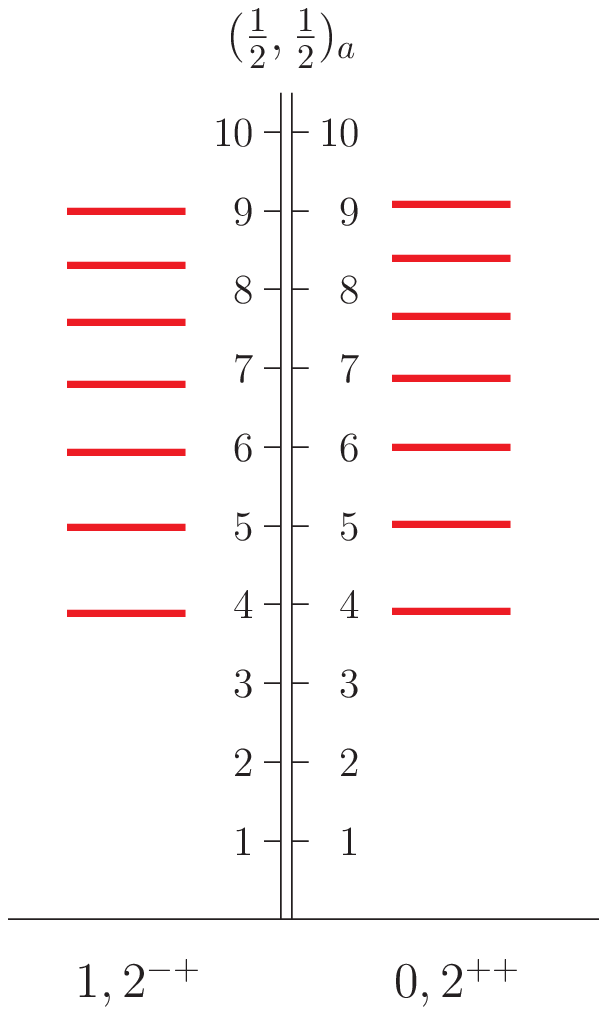}\,%
\includegraphics[width=0.16\hsize]{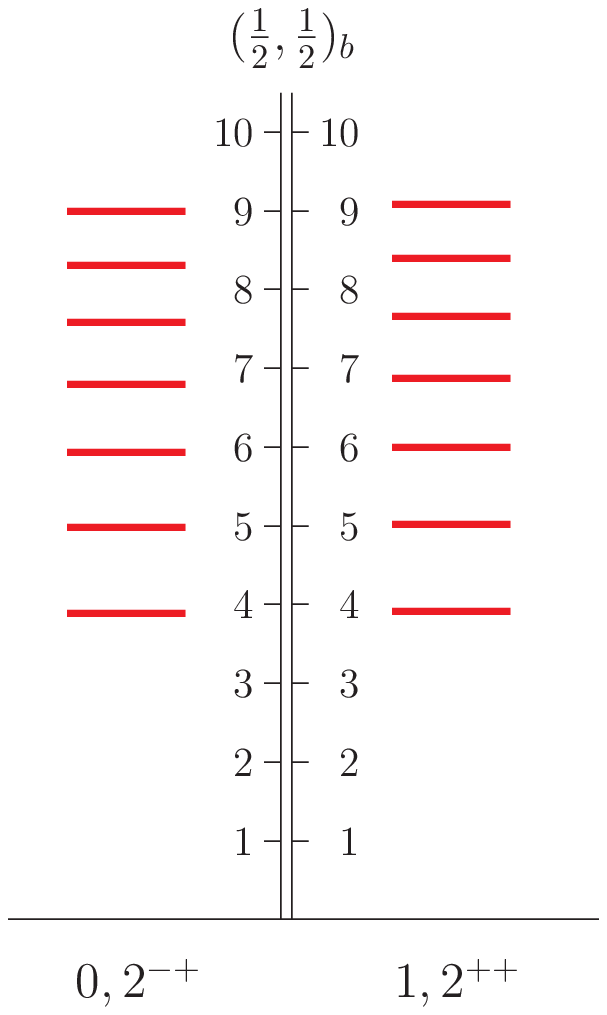}\,%
\includegraphics[width=0.16\hsize]{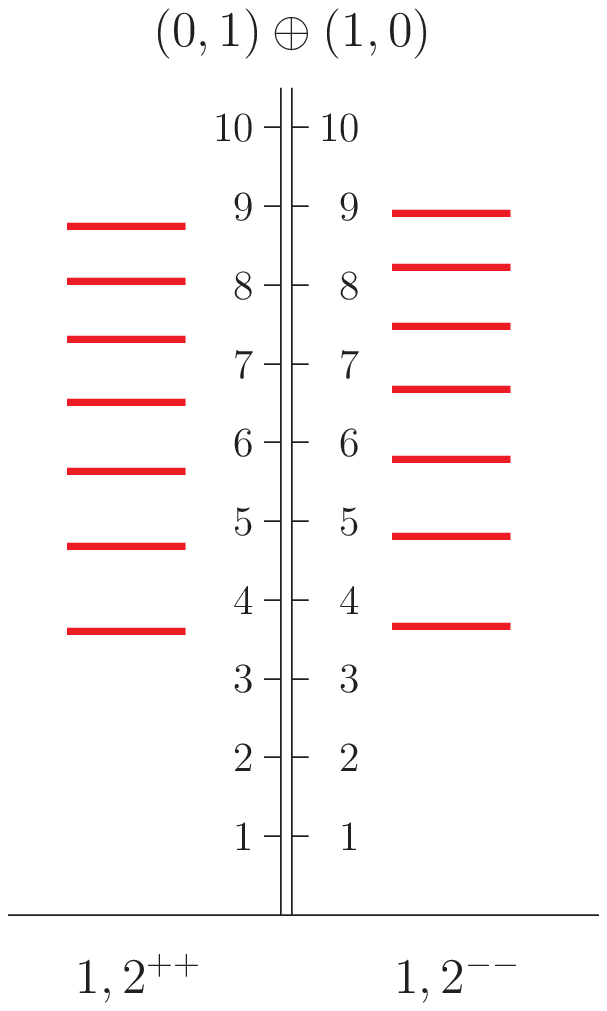}
}
\caption{Spectra for  $p_f=0.05\sqrt{\sigma}$. 
For $J=0$ only the two $(\frac{1}{2},\frac{1}{2})$ multiplets are 
present (left two panels). For $J>0$ there are also the 
$(0,0)$ and $(0,1)\oplus(1,0)$ multiplets. In the remaining four 
panels we show all multiplets for $J=2$. Masses are in units of $\sqrt \sigma$.
Meson quantum numbers are $I,J^{PC}$.}
\end{figure*}
\begin{figure*}
\mbox{
\includegraphics[width=0.16\hsize]{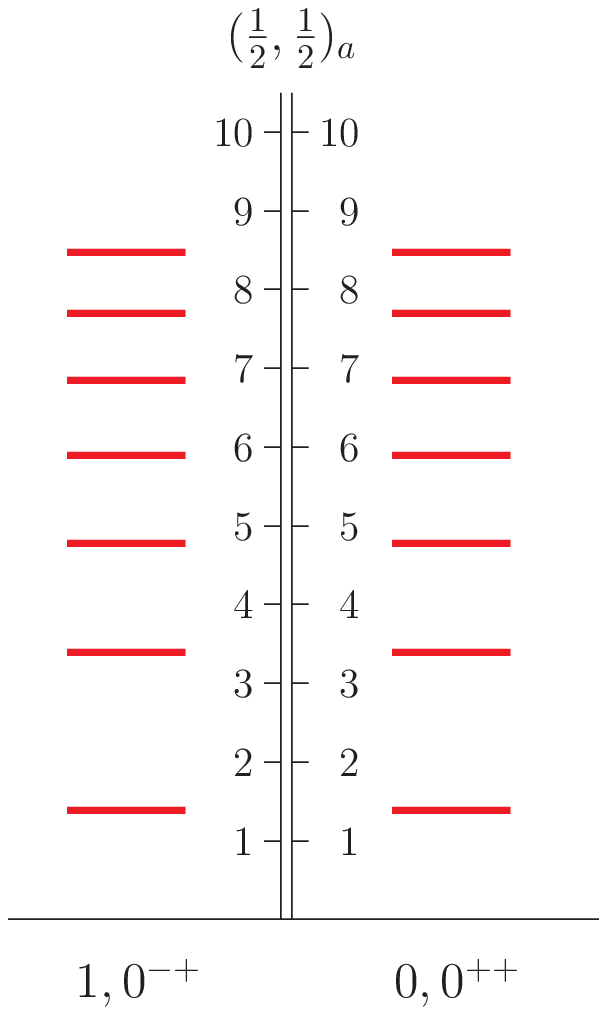}\,%
\includegraphics[width=0.16\hsize]{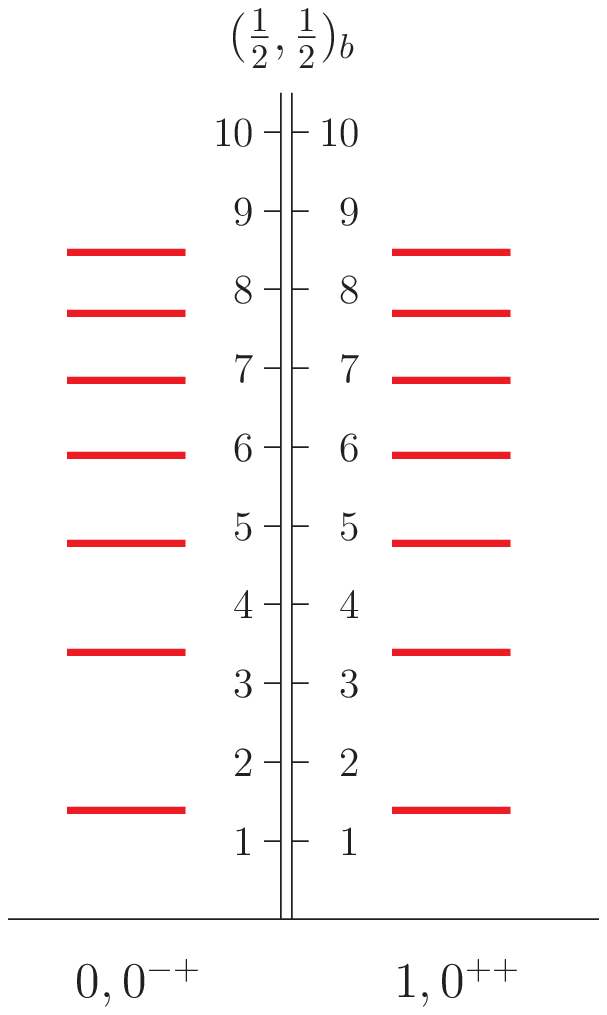}\,%
\includegraphics[width=0.16\hsize]{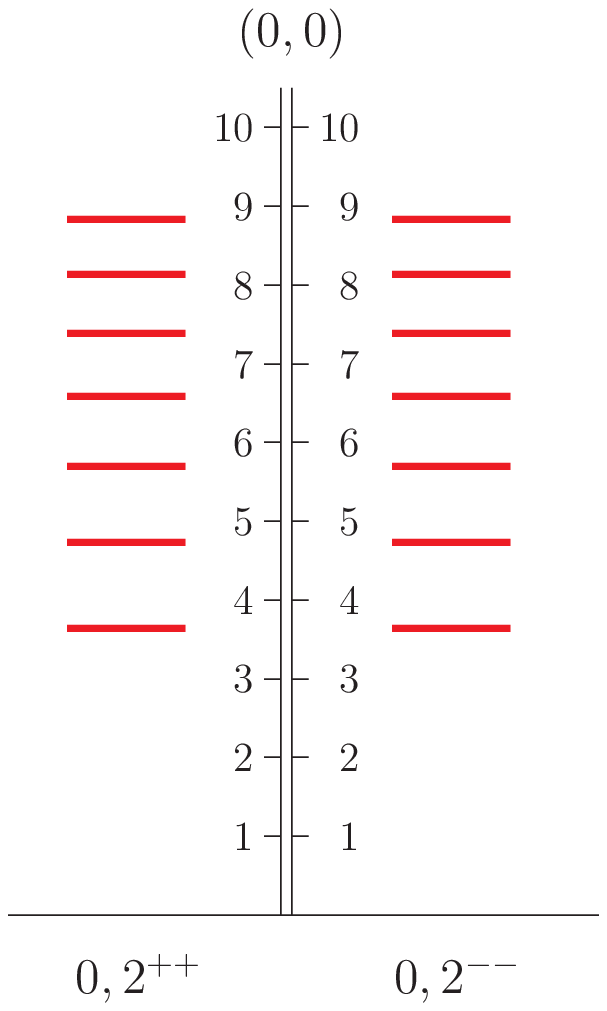}\,%
\includegraphics[width=0.16\hsize]{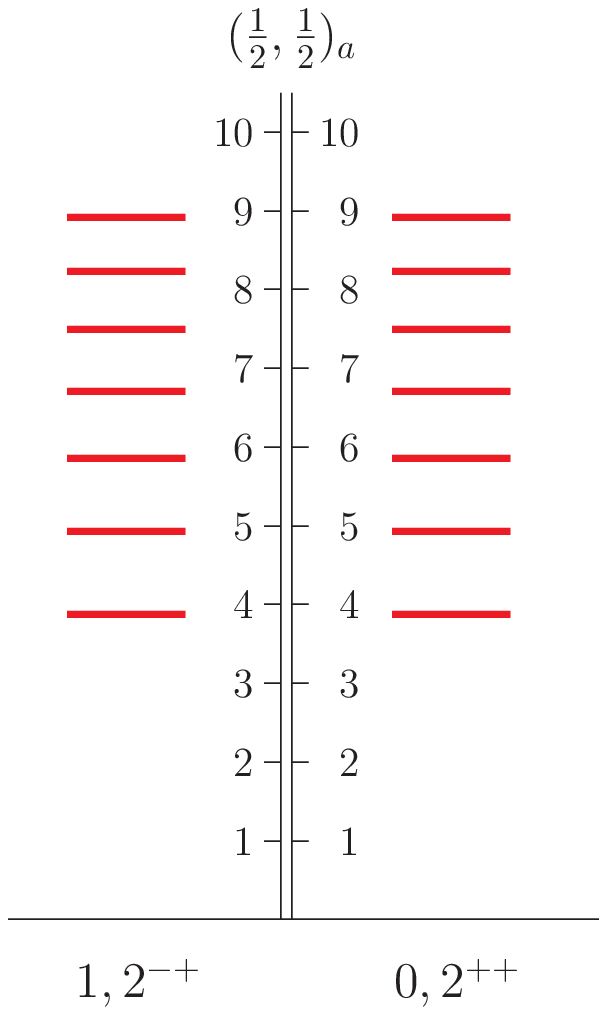}\,%
\includegraphics[width=0.16\hsize]{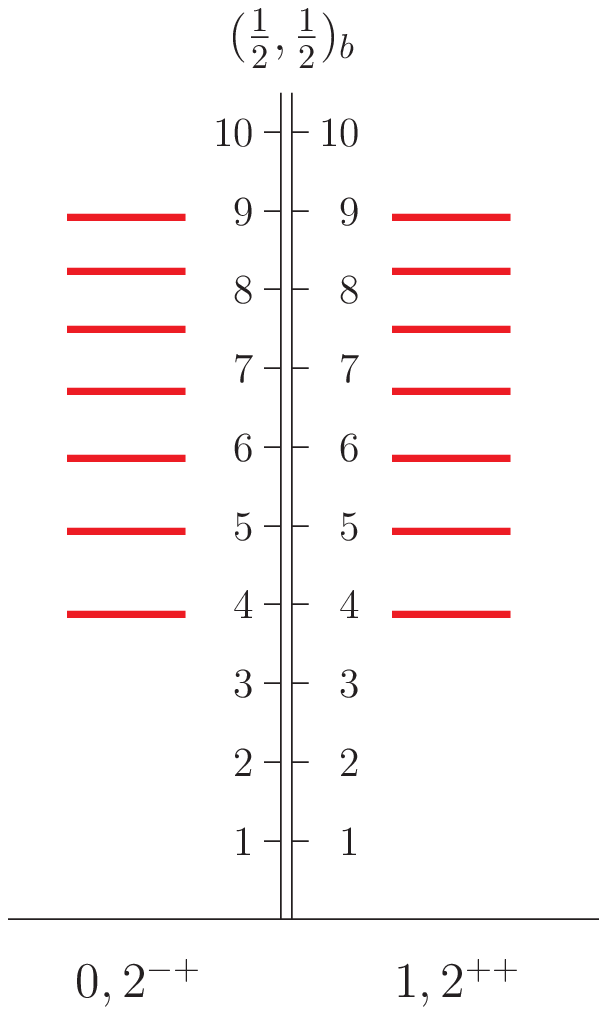}\,%
\includegraphics[width=0.16\hsize]{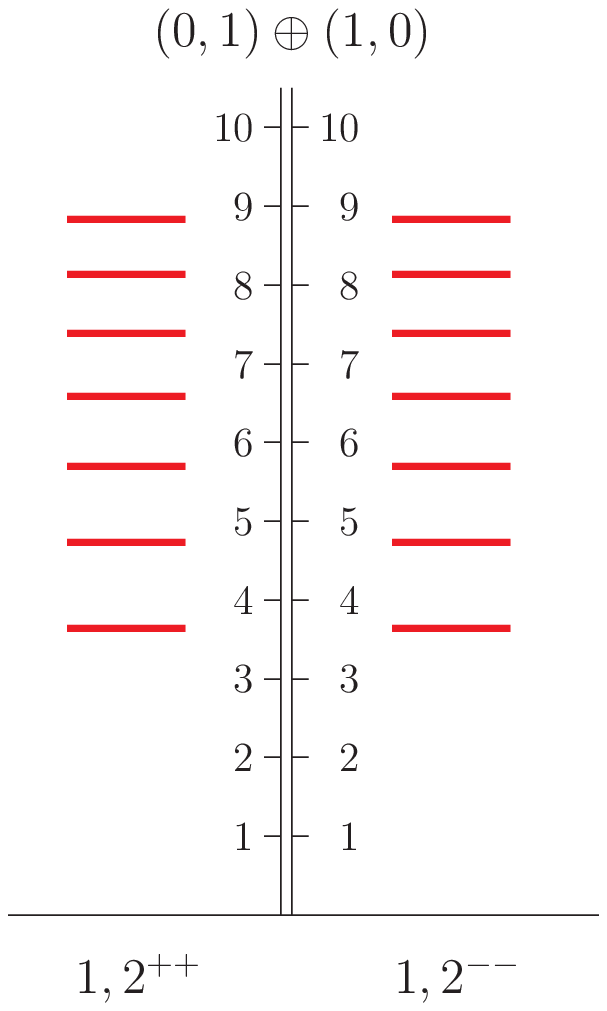}
}
\caption{As Fig. 4 but for Fermi momentum $p_f=0.2\sqrt{\sigma}$.}
\end{figure*}
%\end{widetext}
%

\medskip
{\bf Acknowledgements}
Support of the Austrian Science
Fund through the grant P21970-N16 is acknowledged.

\end{document}